\documentclass[conference]{IEEEtran}
\pdfoutput=1

\IEEEoverridecommandlockouts
\usepackage{cite}
\usepackage{cancel}
\usepackage{float}
\usepackage{amsmath,amssymb,amsfonts}
\usepackage{algorithmic}
\usepackage{subcaption}
\usepackage{graphicx}
\usepackage{textcomp}
\usepackage{xcolor}
\usepackage{url}
\usepackage{algorithm}
\usepackage{balance}

\usepackage[shortlabels]{enumitem}
\def\BibTeX{{\rm B\kern-.05em{\sc i\kern-.025em b}\kern-.08em
    T\kern-.1667em\lower.7ex\hbox{E}\kern-.125emX}}

\def \toolname {FACOS}
\def \threadPair {\textit{thread content}}
\def \threadEmb {\textit{thread embedding}}
\def \apiEmb {\textit{method embedding}}
\def \apiPair {\textit{method content}}
\def \TextCodeEmb {\textit{API relevance embedding}}
\def \TextCodeClassifier {\textit{API relevance classifier}}

\def \datyspScore {DATYS+ score}
\def \relScore {joint relevance score}

\def \clsScoreMath {B}
\def \datyspScoreMath {A}
\def \relScoreMath {C}

\def \fonescore {$F_{\textit{1}}\textit{-}score$}
\def \fonescores {$F_{\textit{1}}\textit{-}scores$}

\def \combineVar {weighting factor}

\def \combineVarMath {x}

\newboolean{showcomments}
\setboolean{showcomments}{true}
\ifthenelse{\boolean{showcomments}}
{ \newcommand{\mynote}[2]{
            \textcolor{blue}{#1}
			\textcolor{red}{
			    \fbox{\bfseries\sffamily\scriptsize David}
			    {\small$\blacktriangleright$\textsf{#2}$\blacktriangleleft$}}}}
{ \newcommand{\mynote}[2]{}}

\ifthenelse{\boolean{showcomments}}
{ \newcommand{\hnote}[2]{\textcolor{red}{
			\fbox{\bfseries\sffamily\scriptsize#1}
			{\small$\blacktriangleright$\textsf{\emph{#2}}$\blacktriangleleft$}}}}
{ \newcommand{\hnote}[2]{}}

\begin{document}

\title{\toolname{}: Finding API Relevant Contents on Stack Overflow with Semantic and Syntactic Analysis \\
}

\author{\IEEEauthorblockN{Kien Luong\IEEEauthorrefmark{1}, 
Mohammad Hadi\IEEEauthorrefmark{2},
Ferdian Thung\IEEEauthorrefmark{1},
Fatemeh Fard\IEEEauthorrefmark{2}, and
David Lo\IEEEauthorrefmark{1}}
% Eldon Tyrell\IEEEauthorrefmark{4},~\IEEEmembership{Fellow,~IEEE}}
\IEEEauthorblockA{\IEEEauthorrefmark{1}School of Computing and Information Systems,
Singapore Management University}%
\IEEEauthorblockA{\IEEEauthorrefmark{2}Irving K. Barber Faculty of Science, University of British Columbia}%
\IEEEauthorblockA{\{kiengialuong, ferdianthung, davidlo\}@smu.edu.sg}%
\IEEEauthorblockA{\{mohammad.hadi, fatemeh.fard\}@ubc.ca}%
}

\maketitle

\begin{abstract}
Collecting API examples, usages, and mentions relevant to a specific API method over discussions on venues such as Stack Overflow is not a trivial problem. It requires efforts to correctly recognize whether the discussion refers to the API method that developers/tools are searching for. The content of the thread, which consists of both text paragraphs describing the involvement of the API method in the discussion and the code snippets containing the API invocation, may refer to the given API method. Leveraging this observation, we develop \toolname{}, a context-specific algorithm to capture the semantic and syntactic information of the paragraphs and code snippets in a discussion. \toolname{} combines a syntactic word-based score with a score from a predictive model fine-tuned from CodeBERT. \toolname{} beats the state-of-the-art approach by 13.9\% in terms of \boldmath{\fonescore{}}.
\end{abstract}

% \begin{IEEEkeywords}
% API searching, Mining
% \end{IEEEkeywords}

\section{Introduction}\label{sec:introduction}
Developers typically use existing libraries or frameworks to implement certain common functionalities. Understanding which APIs to use, the methods they offer, their distinctive names, and how to use them is vital in this regard.
There may be hundreds or even thousands of APIs in a large-scale software library such as the .NET framework and JDK. Microsoft conducted a survey in 2009 in which 67.6\% of respondents said that inadequate or absent resources hindered learning APIs \cite{MS_survey}.

In order to gain a deeper understanding of APIs and their usage information, developers need to inspect many web pages manually and they use automated code search tools.
Most of the Code search tools do not consider the semantics of natural language queries because they are based on keyword matching. 
Stack Overflow %and GitHub 
is the second most common place for the developers to discover APIs, their simple method names, and their usage through crowd-sourced questions and answers. As many API names share simple names but provide different functionality, it is difficult to find code snippets and APIs that correspond to the specific problem searched by the developers on these platforms.
Moreover, API mentions in the informal text content of Stack Overflow are often ambiguous, which makes it difficult to track down APIs and learn their uses. 

Developers frequently discuss and mention APIs in natural language in online discussion and question answering forums like Stack Overflow~\cite{venkatesh2016client, linares2014api, parnin2012crowd, uddin2019automatic}. When developers or automated tools are looking for a specific API, names of API methods sharing the same name can be ambiguous. Therefore, we require API disambiguation to support several downstream tasks such as API recommendation and API mining. To properly index and link APIs to their related information in various sources (e.g., Stack Overflow, Javadoc, etc.), it is important to link ambiguous API mentions to their actual APIs correctly.

Luong et al. recently proposed DATYS~\cite{kien2021datys}, which uses type-scoping to disambiguate API mentions in informal text content on Stack Overflow. In type scoping, they considered API methods whose types appear in more parts (i.e., scopes) of a Stack Overflow thread as more likely to refer to the searched API method. 
However, the statistical word alignment model it uses is based on the appearance of words in a sentence rather than considering in which context the sequence of words are being used and what connotations do these words relaying to the readers. 

APIs are often discussed and mentioned in natural language in online forums such as Stack Overflow to better understand them. 
Developers or automated tools looking for a specific API would be confused by API methods with the same name.
In Stack Overflow, API disambiguation is crucial to finding APIs. Several downstream tasks, such as API recommendation~\cite{huang2018api, rahman2016rack}, are supported by this collection, including API mining~\cite{huang2018api, rahman2016rack}, which relies on interpreting API mentions correctly to index and link APIs to relevant information in various data sources, including Stack Overflow, Javadoc, etc.

To incorporate a deeper understanding of the underlying semantics in a natural language text content of Stack Overflow, we introduce \toolname{}, a context-specific algorithm to capture the semantic and syntactic information of the paragraphs and code snippets in a crowd-sourced discussion. 
We call this API resource retrieval task because \toolname{} focuses on finding Stack Overflow threads mentioning a given API method.
Our work also modified DATYS to perform better search over the code snippets in the Stack Overflow discussion threads. The modified DATYS, denoted as DATYS+ provides an additional metric to better capture the occurrence of API method type in the Stack Overflow thread. By greedily matching the type name with the tokens in the code snippet, DATYS+ performs the syntactic search in \toolname{}. Yet, both DATYS and DATYS+ are only searching based on the syntactic information provided by the fully qualified name of a target API method. It cannot capture the semantic meaning in paragraphs and code snippets of threads on Stack Overflow and how similar they are to the target API method. Thus, to capture the semantics, in addition to the weighted syntactic information provided by DATYS+, \toolname{} has a semantic search component that leverages a deep attention based Transformer model, CodeBERT \cite{feng2020codebert}. This semantic search component measures the similarity between the paragraphs and code snippets of a Stack Overflow thread with the target API method comment and implementation code. The more similar they are, the more likely the thread to be relevant to the target API method that we search for. 
To efficiently leverage both semantic and syntactic knowledge of Stack Overflow thread and the API method, \toolname{} joined the semantic and the syntactic search element to get the \relScore{} that determines whether a thread relates to a given API method. The contributions of each element in the \relScore{} is defined by a \combineVar{}.

In this paper, we are going to answer with the following research questions:

\begin{enumerate}[leftmargin=25pt]
  \item[\textbf{RQ1}] Can \toolname{} perform better than the baseline (DATYS)?
  \item[\textbf{RQ2}] How well does each component of \toolname{} perform?
  \item[\textbf{RQ3}] How does the \combineVar{} affect the F1-score of \toolname{}?
\end{enumerate}

These research questions will help us understand the effectiveness of our approach \toolname{} and the internal mechanism through which it yields better results than the current baseline.
Our work has offered the following main contributions: 

\begin{enumerate}
    \item To our knowledge, we are the first to adopt a transformer-based deep learning technique to incorporate semantic knowledge understanding for API resource retrieval task. 
    \item As compared to state-of-the-art techniques, our approach performs better while searching for the contents related to the queried API. On a dataset of 380 Stack Overflow threads, \toolname{} beats the state-of-the-art by 13.9\%.
    \item We designed an ablation study to understand how the integrated components of our approach are performing. We found that each component contributes to the effectiveness of \toolname{}.
    \item We have also open sourced our code and additional artifacts required for recreating the results and re-purposing our approach for other tasks. The source code for \toolname{} is available at \url{https://anonymous.4open.science/r/facos-E5C6/}
\end{enumerate}

The rest of the paper is structured as follows: Section \ref{sec:preliminary} deals with the preliminary knowledge about the components on top of which we have built our method. Section \ref{sec:approach} 
% details the devised methodology and section 
provides an overview of our proposed approach, while Section
\ref{sec:fusion} 
% describes the designed model and its components. 
elaborates the various components of our proposed approach.
We 
% have detailed our experiments 
describe our experiment details
and results in Section \ref{sec:experiment} and \ref{sec:result}, respectively. The related works and the threats to validity are presented in Sections \ref{sec:related_works} and \ref{sec:threats_to_validity}. Finally, we concluded our work and present future work in Section \ref{sec:conclusion}.

\section{Preliminaries}\label{sec:preliminary}

\subsection{DATYS}\label{subsec:datys}
Two steps are involved in finding the API mentioned in informal text content: (1) API mentions extraction, and (2) API mentions disambiguation. API mention extraction aims to identify common words that refer to the APIs. API mentions disambiguation, on the other hand, links API mentions with the APIs they reference. DATYS \cite{kien2021datys} specifically deals with the API mention disambiguation via type scoping in the informal text of Stack Overflow to resolve ambiguous mentions of Java API methods, where the mentions have been identified. 

After extracting API method candidates from input Java libraries, DATYS scores API method candidates based on how often their types (i.e., classes or interfaces) appear in different parts (i.e., scopes) of the Stack Overflow thread with identified API mentions. Having a type that appears in more scopes will increase the API candidate score. Here, DATYS considers three scopes:
{\em Mention scope}, which covers the mention itself.
{\em Text scope}, which covers the textual content of the thread, including the mentions.
{\em Code scope}, which covers the code snippets in the thread.
API candidates are ranked according to their scores for each API mention in the thread. DATYS takes the top API candidate with a non-zero score as the mentioned API. If the leading API candidate has a zero score, DATYS considers the mention as an unknown API. Luong {\em et al.} built a ground truth dataset containing 807 Java API mentions from 380 threads in Stack Overflow.

\subsection{CodeBERT} \label{subsec:codebert}
CodeBERT \cite{feng2020codebert} was developed using a multilayered attention-based Transformer model, BERT \cite{devlin2018bert}. As a result of its effectiveness in learning contextual representation from massive unlabeled text with self-supervised objectives, the BERT model has been adopted widely to develop large pre-trained models.
Thanks to the multilayer Transformer~\cite{vaswani2017attention},
CodeBERT developers adopted two different approaches than BERT to learn semantic connections between Natural Language (NL) - Programming Language (PL) more effectively.

Firstly, The CodeBERT developers make use of both bimodal instances of NL-PL pairs (i.e., code snippets and function-level comments or documentations) and a large amount of available unimodal codes. In addition, the developers have pre-trained CodeBERT using a hybrid objective function, which includes masked language modeling~\cite{devlin2018bert} and replaced token detection~\cite{clark2020electra}. The incorporation of unimodal codes helps the replaced token detection task, which in turn produces better natural language by detecting plausible alternatives sampled from generators.

Developers trained CodeBERT from Github code repositories in 6 programming languages, where only one pre-trained model is learned for all six programming languages with no explicit indicators used to mark an instance to the one out of six input programming languages. 
CodeBERT was evaluated on two downstream tasks: natural language code search and code documentation generation. The study found that fine-tuning the parameters of CodeBERT obtained state-of-the-art results on both tasks.

\section{Approach Overview}\label{sec:approach}
\subsection{Task Definition}
Our goal is to find Stack Overflow threads that mention a given API method\footnote{In this paper, we use the terms API, method, and API method interchangeably.}. 
Specifically, given an API method, we strive to find Stack Overflow threads containing words matching the simple name of the given API method. In Java, the simple name of an API method is the name of the method without the class and the package names. For example, {\tt m} is the simple name of API method {\tt com.example.Class.m}. We want to classify whether the threads having the simple name {\tt m} is actually relevant to API method {\tt com.example.Class.m}.  
In summary, the task is defined as: \textit{``For each API method in a set of given API methods, identify Stack Overflow threads that refer to it.''}

\subsection{Architecture}\label{subsec:architecture}
\begin{figure}[t]
\centerline{\includegraphics[width=\columnwidth]{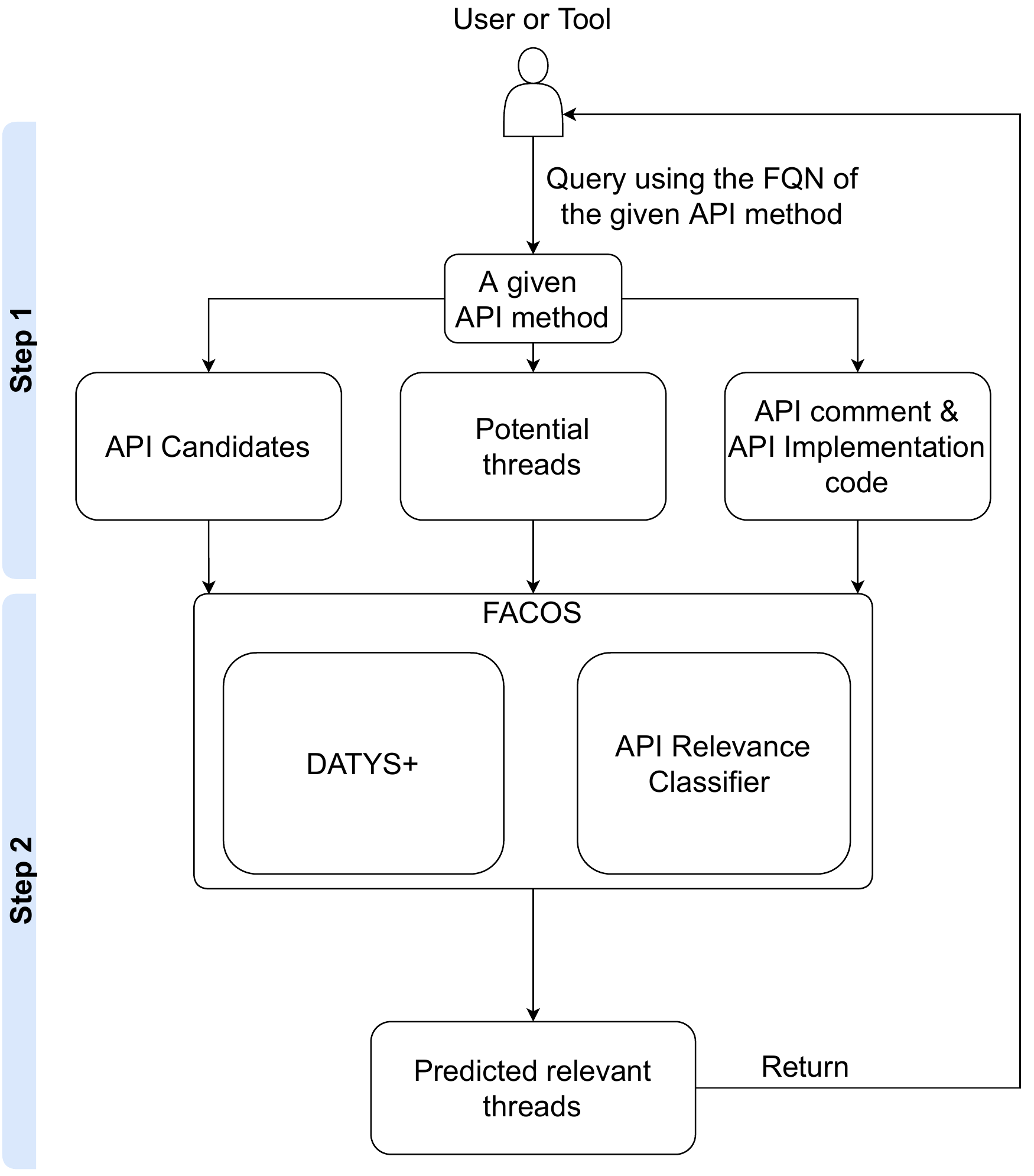}}
\caption{The architecture of \toolname{} (\underline{F}inding \underline{A}PI Relevant \underline{C}ontents on Stack \underline{O}verflow with \underline{S}emantic and Syntactic Analysis)}
\label{fig:overview}
\end{figure}

The pipeline of \toolname{} is presented in Figure~\ref{fig:overview}.
It is divided into 2 main steps:

(1) Collecting various API-related resources from a given API method name; 
and (2) Recommending relevant threads using the collected API-related resources. 

In the step (1), \toolname{} finds {\em Potential Threads} from Stack Overflow using the simple name of the given API method as the query. {\em Potential Threads} are the threads that have at least one word matching with the simple name of given API. 
The {\em API method comment and implementation code} are directly obtained from the source code repository of the given API. Last but not least, the {\em API Candidates} are obtained from a database of API methods. The {\em API Candidates} are API methods that have the same simple name as the given API method.

The objective of step (2) is to identify whether each Stack Overflow thread in the {\em Potential Threads} actually refers to the given API. \toolname{} has two components: {\em \TextCodeClassifier{}} and DATYS+. {\em \TextCodeClassifier{}} is designed to draw the relevance between a thread and an API method by capturing the semantic similarity between (1) paragraphs and code snippets in the thread; and (2) API method comment and implementation code. {\em \TextCodeClassifier{}} outputs a semantic relevance score representing the relevance it measures.
In contrast, DATYS+ outputs a syntactic relevance score based on the existence of the terms from the fully qualified name of the given API in different scopes of a Stack Overflow thread. For example, API "A.B.C" has terms such as "A", "B", and "C". The last term, "C", is the simple name of the API method. The second last term, "B", is the type of the API method. 
Both DATYS and DATYS+ use type scoping~\cite{kien2021datys} to give score based on the existence of the type of the API (i.e., "B" in the example) in different scopes of the Stack Overflow thread (code scope, text scope, etc.). 
However, type scoping of DATYS+ is modified to be suitable for the search task and we are going to describe it in Section~\ref{subsec:datysp}. It outputs a score that indicates the syntactic relevance between the given API and the thread, we call it DATYS+ score.

After step (2), each thread will have a score indicating if the thread refers to the given API method. This score is combined from semantic relevance score and DATYS+ score and is called {\em \relScore{}}. Threads predicted as referred to the given API method are then returned to the user. We describe \toolname{} components (i.e., DATYS+ and \TextCodeClassifier{}) in detail in Section~\ref{sec:fusion}.

\section{\toolname{}}\label{sec:fusion}
\toolname{} consists of two main components: DATYS+ and \TextCodeClassifier{}. DATYS+ takes as inputs {\em Potential Threads} and {\em API Candidates} and outputs scores indicating its confidence that the given API is referred to in the threads (Section~\ref{subsec:datysp}). Given {\em Potential Threads} and {\em API method comment and implementation code}, \toolname{} first converts them to \TextCodeEmb{} (Section~\ref{subsec:relEmb}). The \TextCodeEmb{} is input to \TextCodeClassifier{}, which outputs confidence scores indicating the likelihood that the given threads refer to the API (Section~\ref{subsec:relclassifier}). Finally, the scores from DATYS+ and \TextCodeClassifier{} are combined to a \relScore{}, and threads with scores larger than a threshold are returned as the relevant threads (Section~\ref{subsec:relscore}).

\subsection{DATYS+}\label{subsec:datysp}
DATYS+ is an extension of DATYS.  DATYS used regular expressions to capture the types of API method invocations available in code snippets of the thread. However, these regular expressions are limited and thus DATYS may miss some mentions in code snippets. To capture more types, DATYS+ modifies the type scoping algorithm by adding a new score.

Algorithm~\ref{algo:typescoping} indicates how modified type scoping works. % for the task of API relevant content search.
Compared to DATYS's, DATYS+'s type scoping algorithm receives $\mathit{CodeSnippets}$ as another input.
$\mathit{CodeSnippets}$ represents the content available in code snippets of the Stack Overflow thread. In addition, inputs of the original type scoping algorithm are also considered. $\mathit{APIMention}$, $\mathit{PTypeList}$, $\mathit{APIMethodCandidate}$, and $\mathit{ThreadContent}$ stand for the simple name of the given API, the list of possible types extracted from code snippets following the algorithm used by DATYS, the {\em API Candidates}, and the thread's textual content (i.e., title, text, tags), respectively.
The three scopes used by DATYS are also used in DATYS+. In \textit{Mention Scope (Lines~\ref{algo:mentionStart}-\ref{algo:mentionEnd})}, DATYS+ increases an API score if its type appear within the API mention. In \textit{Text Scope (Lines~\ref{algo:textTokenStart}-\ref{algo:textTokenEnd})}, DATYS+ increases an API score if its type appear within the textual content of the thread. In \textit{Code Scope (Lines~\ref{algo:codeStart}-\ref{algo:codeEnd})}, DATYS+ increases an API score if its type matches with the type of method invocation or imported types in the code snippet. Additionally, in \textit{Code Scope}, DATYS+ also looks at the content of the code snippets and increases the API score of the corresponding API candidate if there are tokens in the code snippets that match with the API type (Lines~\ref{algo:codeTokenStart}-\ref{algo:codeTokenEnd}).
This score helps to capture the occurrence of types that would be missed by a more accurate matching used in DATYS. Thus, we call the scope of this score {\em Extended Code Scope}

After executing type scoping, DATYS+ returns scores for the {\em API Candidates}. The scores are then normalized to a range of [0, 1] following the minimum and the maximum score from the {\em API Candidates}. DATYS+ then takes the normalized score of the given API method and passes it to the next step.

\begin{algorithm}[t]
\caption{Scoring an API Candidate with Type Scoping in DATYS+}\label{algo:typescoping}
\footnotesize{
\begin{algorithmic}[1]
\renewcommand{\algorithmicrequire}{\textbf{Input:}}
\renewcommand{\algorithmicensure}{\textbf{Output:}}
\REQUIRE \textit{$ApiMention, PTypesList, APIMethodCandidate,$\\$ThreadContent, CodeSnippets$}
\ENSURE  \textit{$CandScore$}
\\% \textit{Initialisation} :
\STATE $CandScore$ = 0 \label{algo:init}
\STATE $CandType = getType(APIMethodCandidate)$ \label{algo:getType}

\IF {$hasPrefix(ApiMention)$} \label{algo:mentionStart}
    \STATE \textit{$Prefix = getPrefix(ApiMention)$}
    \IF {$endsWith(Prefix,CandType)$}
        \STATE \textit{$CandScore = CandScore + 1$}
        
    \ENDIF
\ENDIF \label{algo:mentionEnd}
\STATE $TextualTokens = tokenize(ThreadContent)$
\STATE $CodeTokens = tokenize(CodeSnippets)$
\label{algo:textTokenStart}
\IF {$CandType$ $in$ $TextualTokens$}
    \STATE \textit{$CandScore = CandScore + 1$}
\ENDIF \label{algo:textTokenEnd}

\IF {$CandType$ $in$ $CodeTokens$}\label{algo:codeTokenStart}
    \STATE \textit{$CandScore = CandScore + 1$}
\ENDIF \label{algo:codeTokenEnd}
% \\ \textit{LOOP Process}
\FOR {$PType$ in $PTypesList$} \label{algo:codeStart}
% \STATE statements..
\IF {$isSameType(PType, CandType)$ }
    \STATE \textit{$CandScore = CandScore + 1$}
    
    % \IF {($isSubString(PType, ParagraphOfMention)$) }
    % \STATE \textit{$CandidateScore = CandidateScore + 1$}
    % \ENDIF
   
\ENDIF
\ENDFOR \label{algo:codeEnd}

\RETURN $CandScore$
\end{algorithmic}
}
\end{algorithm}

\subsection{\TextCodeEmb{}}\label{subsec:relEmb}
\begin{figure}[t]
\centerline{\includegraphics[width=\columnwidth]{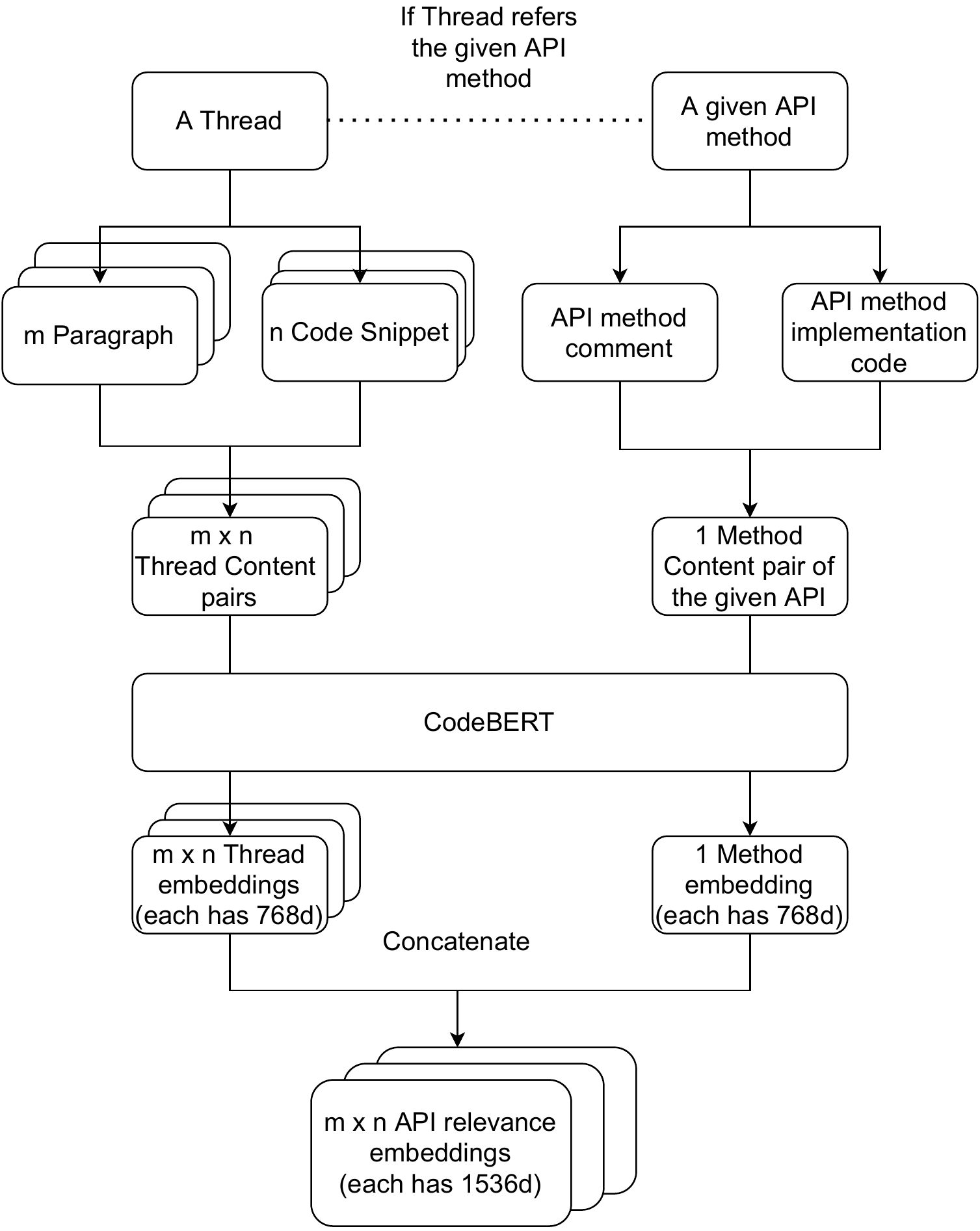}}
\caption{How \TextCodeEmb{}s are created}
\label{fig:text_code_emb}
\end{figure}

We follow the process described in Figure~\ref{fig:text_code_emb} to build \TextCodeEmb{}.
Firstly, each thread in {\em Potential Threads} needs to be converted into an embedding.
A thread may contain {\em m} paragraphs and {\em n} code snippets. 
A paragraph is a piece of textual content on a Stack Overflow thread that is separated from other contents in the thread via a newline character. %, it does not start with any xml tag for example 
Code snippet is a piece of code content on a Stack Overflow thread. It is typically enclosed with a starting tag $ \langle pre \rangle\langle code \rangle$ and an ending tag  $\langle /code \rangle\langle /pre \rangle$. %</code></pre>
Each paragraph is paired with each code snippet to create a pair of \threadPair{}. Therefore, a Stack Overflow thread would have $m\times n$ \threadPair{} pairs. A natural-programming language model, CodeBERT\footnote{\url{https:// github.com/microsoft/CodeBERT}}, is used to extract the semantic meaning of each \threadPair{} pair. It encodes the  $m\times n$ \threadPair{} pairs into $m\times n$ \threadEmb{}s. 
\threadEmb{} is the representation vector of \threadPair{} that created by CodeBERT's encoder.
By converting the pairs from a textual form to a numerical vector form with a pre-trained CodeBERT model, the semantic relationship between the paragraphs and code snippets is extracted.
Before feeding the \threadPair{} pairs into the encoder of CodeBERT, each pair is pre-processed following the format:
\[
\langle CLS \rangle\ paragraph\ \langle SEP \rangle\ code\ snippet\ \langle EOS \rangle
\]

\noindent $\langle CLS \rangle$ is the token that informs the start of the pair according to the design of RoBERTa model~\cite{liu2019roberta} which CodeBERT is based on.  $\langle SEP \rangle$ is the token that separates a Paragraph from a Code Snippet and $\langle EOS \rangle$ indicates the end of the pair. 
In detail, the maximum number of tokens in a pair before being fed into CodeBERT encoder is 512.
We set the number of tokens for a paragraph and a code snippet to 254 and 255 tokens, respectively. The two numbers add up to 512 when the three tokens such as $\langle CLS \rangle$, $\langle SEP \rangle$, and $\langle EOS \rangle$ are counted.
If the number of tokens in the paragraph is less than 254, then padding tokens would be added to reach 254 tokens. On the other hand, if the number of tokens in the paragraph is more than 254, we truncate the paragraph and take the first 254 tokens. The same process is applied to the code snippet with 255 tokens. The CodeBERT encoder receives these \threadPair{} pairs under this format as inputs and outputs embedding vectors. For a thread with $m\times n$ \threadPair{} pairs, there would be $m\times n$  \threadEmb{} vectors created and each \threadEmb{} vector has a length of 768.

Secondly, to build \TextCodeEmb{}, {\em API comment and implementation code} also need to be converted into an embedding.
The API method comment is a piece of textual content that describes the functionality of the API method and how to use it. The API implementation code is the code inside the API method body that implements the described functionality.
The {\em API comment and implementation code} are extracted from the Javadoc and the JAR files, respectively, they are pre-processed to the following format:
\[
\langle CLS \rangle\ comment\ \langle SEP \rangle\ implementation\ code\ \langle EOS \rangle
\]
They are then transformed into a numerical representation vector via the CodeBERT encoder.

Finally, each \threadEmb{} vector and the \apiEmb{} vector are then concatenated to a vector. We call this concatenated vector \TextCodeEmb{}. In total, $m\times n$ \TextCodeEmb{} vectors would be created.

\subsection{\TextCodeClassifier{}}\label{subsec:relclassifier}
The \TextCodeClassifier{} is a binary classifier that utilizes a neural network with two fully connected layers to predict whether the \TextCodeEmb{} comes from a Stack Overflow thread that refers to the given API method.

The \TextCodeClassifier{} has two modes of operation: training and deployment modes. In the training mode, the \TextCodeEmb{s} are used to train the \TextCodeClassifier{}. When there is an imbalance between positive and negative labels, \TextCodeClassifier{} upsamples the minority label.
Whenever the thread refers to the given API method, all \TextCodeEmb{} created from the thread would be considered as positive by the classifier. Otherwise, in case the given API method is not referred to by the thread, every \TextCodeEmb{} of the thread would have negative labels.
In the deployment mode, \TextCodeClassifier{} produces probability scores for the $m\times n$ \TextCodeEmb{}. These scores are averaged and passed to the next step. The averaged score indicates the likelihood that the thread refers to the given API.

\subsection{Computing joint relevance score}\label{subsec:relscore}
\begin{figure}[t]
\centerline{\includegraphics[width=\columnwidth]{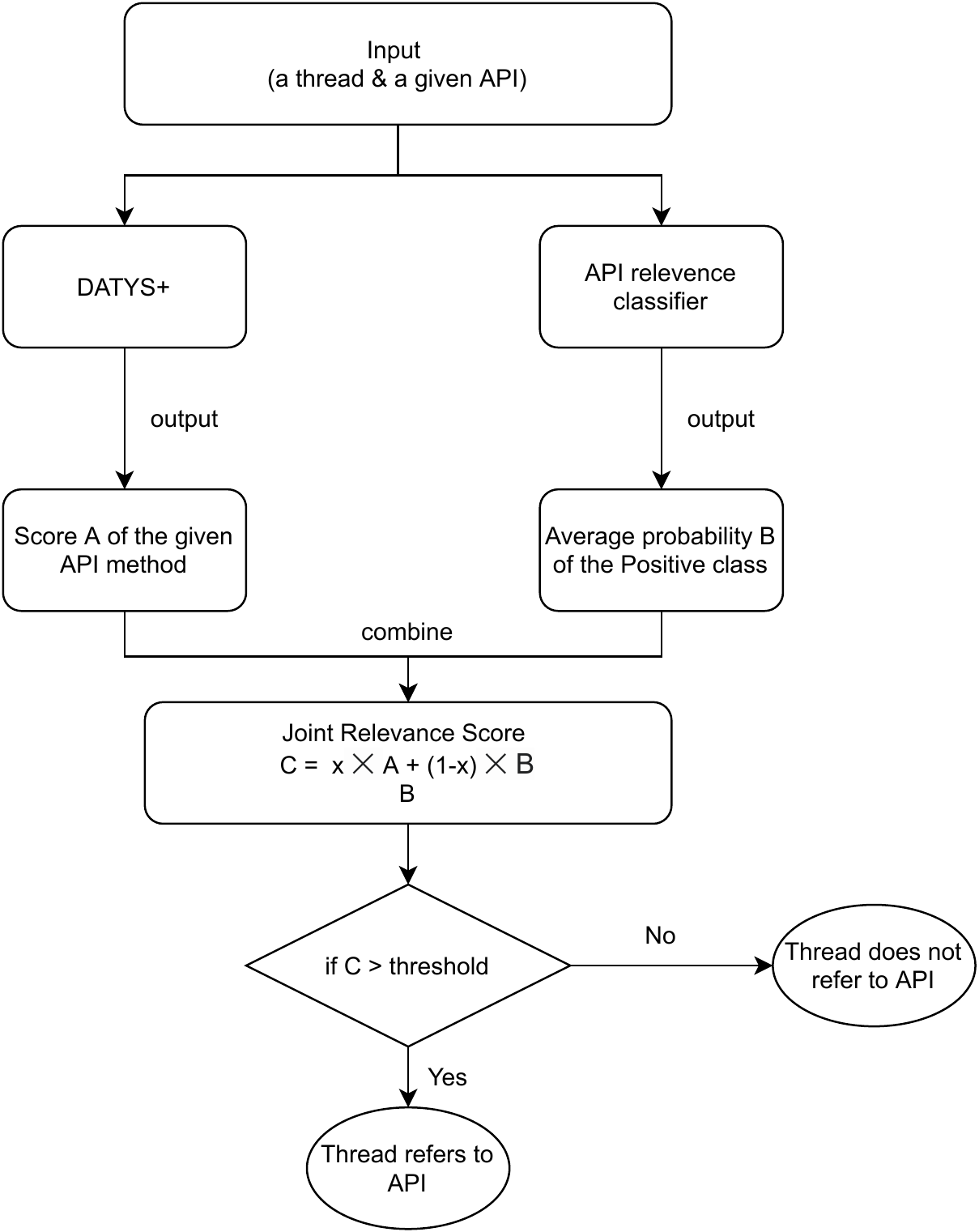}}
\caption{Computing joint relevance score}
\label{fig:relevance_score}
\end{figure}

We follow the process in Figure~\ref{fig:relevance_score} to compute the joint relevance score. DATYS+ and \TextCodeClassifier{} output scores $\datyspScoreMath{}$ and $\clsScoreMath{}$, respectively. Both represent their confidence that the given API method is mentioned in the thread.
The two scores are then combined to a joint relevance score $C$ following this formula:

\begin{equation}
\relScoreMath{} =  x \times \datyspScoreMath{} + (1-x) \times \clsScoreMath{}
\end{equation}

\noindent The \combineVar{} $\combineVarMath{}$ decides the contributions of \datyspScore{} and \TextCodeClassifier{} in \relScore{} $C$. The higher the value of $\combineVarMath{}$ is, the more \datyspScore{} contributes to the final \relScore{}. 
The range of $\datyspScoreMath{}$, $\clsScoreMath{}$, and $\combineVarMath{}$ is from $0$ to $1$.
A thread is considered to refer to the given API if the joint relevance score $C$ is larger than a threshold $t$. Otherwise, the thread is considered not to refer to the given API. By default, $t$ is set to $0.5$.

The value of $\combineVarMath{}$ will be estimated based on the training data. In detail, we let $\combineVarMath{}$ increase gradually from $0$ to $1$ with a step of 0.1. There are ten possible values of $\combineVarMath{}$: $\{0, 0.1, 0.2, ..., 0.9, 1.0\}$.
The value of $\combineVarMath{}$ giving the highest performance in the training data is then chosen.

\section{Experiment}\label{sec:experiment}

\subsection{Dataset and Experimental Settings}
We utilize the dataset provided in DATYS work~\cite{kien2021datys} to evaluate both \toolname{} and DATYS. 
We split 380 Stack Overflow threads to 253 training threads and 127 testing threads with the ratio of 2:1. The training threads are utilized to train the \TextCodeClassifier{} while the testing threads are used to evaluate \toolname{} and DATYS.
Next, as mentioned in Section~\ref{subsec:relEmb}, for each Stack Overflow thread in the training threads, we extract its \threadEmb{}s and these \threadEmb{}s are grouped into a training set. Similarly, for each Stack Overflow thread in the testing threads, we extract its \threadEmb{}s and these \threadEmb{}s are grouped into a testing set.
The numbers of \TextCodeEmb{}s of the dataset are shown in Table~\ref{table:nrofTrainTest}.
The numbers of the embeddings in training set and testing set are $57\,690$ and $26\,212$, respectively.

\begin{table}[t]
    \caption{Number of \TextCodeEmb{}s in each set}
    \centering 
    \begin{center}
        \begin{tabular}{|c|c|}
            \hline
            \textit{}& \textbf{\textit{\# \TextCodeEmb{s}}}\\%& \textbf{\textit{Fold 2}}& \textbf{Fold 3} \\
            \hline
            Training set & 57,690 \\%& 52,789 & 57,640 \\
            \hline
            Testing set & 26,212 \\%& 31,113 & 26,262 \\
            \hline
            \end{tabular}
        \label{table:nrofTrainTest}
    \end{center}
\end{table}

To generate \TextCodeEmb{}s for the \TextCodeClassifier{} for training, for each thread, if the given API appears in the thread, we generate \TextCodeEmb{}s for \threadPair{}s and \apiPair{}s as described in Section~\ref{subsec:relEmb}. These embeddings would have positive label because they are created from the API that is referred to by the thread. To generate embeddings with a negative label for a thread, we find APIs that have the same simple name as the given API and are not mentioned in the thread. We then create \TextCodeEmb{}s from these APIs and label these \TextCodeEmb{}s as negative. 

To train the \TextCodeClassifier{}, there are $344$ APIs. These APIs are used to generate the training \apiEmb{}s. In the testing set, there are $181$ APIs. These APIs are used to generate the testing \apiEmb{}s.
Table~\ref{table:nrofTCEmbs} shows the numbers of positive and negative \TextCodeEmb{}s created in training and testing sets. The number of negative \TextCodeEmb{}s is approximately 4 times more compared to the positive ones in the same set of thread.
Due to this imbalance, positive \TextCodeEmb{}s are randomly up-sampled to balance the two classes within the \TextCodeClassifier{} training process.

\begin{table}[t]
    \caption{Number of positive and negative \TextCodeEmb{}s in each set}
    \begin{center}
        \begin{tabular}{|c|c|}
            \hline
            Positive Embeddings in Training set & 9,934 \\%& 10,555 \\
            \hline
            Negative Embeddings in Training set  & 47,756 \\%&  40,606\\
            \hline
            Positive Embeddings in Testing set & 5,607 \\%& 4,986 \\
            \hline
            Negative Embeddings in Testing set & 20,605 \\%& 27,755 \\
            \hline
            \end{tabular}
        \label{table:nrofTCEmbs}
    \end{center}
\end{table}

The \TextCodeClassifier{} is trained using 6 epochs on the training data. After the first 6 epochs, the value of the loss function has relatively converged. The learning rate of the training is set to $10^{-3}$.

\subsection{Metrics}\label{subsec:metrics}
To evaluate the proposed approach on identifying threads that are relevant to an API, we use three metrics: Precision, Recall, and \fonescore{}.
In order to calculate the three aforementioned metrics, True Positive, False Positive, and False Negative should be defined first. Our task focuses on finding threads that actually refer to a given API. True Positive is the case where a thread is deemed to be relevant by the approach is indeed relevant. False Positive is the case where the thread that is deemed to be relevant by the approach is actually irrelevant. False Negative is the case where a threads is deemed to be irrelevant by the approach is actually relevant.
The metrics are calculated using the following formulas:
\begin{equation}
Precision = \frac{True\ Positive}{True\ Positive+False\ Positive}
\end{equation}
\begin{equation}
Recall = \frac{True\ Positive}{True\ Positive+False\ Negative}
\end{equation}
\begin{equation}
F_{\textit{1}}\textit{-}score = \frac{2 \times Precision \times Recall}{Precision+Recall}
\end{equation}

\noindent We measure the above scores of all given APIs in the testing set and report the averages of the scores.

\subsection{Research Questions}
\vspace{0.2cm} \noindent \textbf{Research Question 1}: Can \toolname{} perform better than the baseline (DATYS)?

\noindent The baseline, DATYS, was designed for a task of API mention disambiguation. We adopt it to our task of finding threads that are relevant to an API. If DATYS finds an API is mentioned in the thread, the thread is considered to be relevant to the API. To evaluate the improvement that \toolname{} over DATYS, we evaluate them in the testing data set and compare them in terms of \fonescore{}. We also analyse some cases that \toolname{} can resolve and DATYS can not in Section~\ref{subsec:outperformingCases}.

\vspace{0.2cm} \noindent \textbf{Research Question 2}: How well does each component of \toolname{} perform?

\noindent There are three possible variants of of \toolname{} depending on which component that comes along with it. The variants are (1) \toolname{} with \TextCodeClassifier{}; (2) \toolname{} with DATYS+; and (3) \toolname{} with DATYS+ and \TextCodeClassifier{}. \TextCodeClassifier{} is a semantic-based algorithm while DATYS+ is a syntactic-based algorithm.
In this study, we aim to analyze the contribution of each component in \toolname{}.
From the analysis, we would like to answer the question whether combining a semantic-based algorithm and a syntactic-based algorithm leads to a better result than running them individually.

\vspace{0.2cm} \noindent\textbf{Research Question 3}: How does the \combineVar{} affect the F1-score of the relevant thread classification? Does our strategy work well?

\noindent The \combineVar{} is an importance factor that would affect how well \toolname{} perform. We select the importance factor based on the best performance in the training data. We analyze whether our strategy leads to the best performance in the testing data. We vary the values of \combineVar{} in both the training data and the testing data. The values that we use are $\{0, 0.1, 0.2, ..., 0.9, 1.0\}$. We analyze whether picking values in the training data that leads to the best performance in the training data also leads to the best performance in the testing data.

\section{Result}\label{sec:result}

\subsection {RQ1: \toolname{} Effectiveness}

\begin{table}[t]
    \caption{\toolname{} vs DATYS in terms of \boldmath{$F_{\textit{1}}$}-score in the testing set}
    \begin{center}
        \begin{tabular}{|c|c|c|c|}
            \hline
            
            \textit{\bf Approach}& \textit{\bf Avg.} & \textit{\bf Avg.} & \textit{\bf Avg.}\\
            \textit{} & \textit{\bf Precision}& \textit{\bf Recall}& \textit{\boldmath{$F_{\textit{1}}$}\bf -score}\\
            \hline
            \textit{DATYS} & 0.7441 & 0.7703 & 0.7340\\
            \hline
            \textit{\toolname{}} & 0.8697 & 0.9016 & 0.8730\\
            \hline
            \end{tabular}
        \label{table:benchmarkTest}
    \end{center}
\end{table}

\begin{table}[t]
    \caption{Contribution of \toolname{} Components}
    \begin{center}
        \begin{tabular}{|l|l|l|l|}
            \hline
            \textit{\bf Components}& \textit{\bf Avg.} & \textit{\bf Avg.} & \textit{\bf Avg.} \\
            \textit{} & \textit{\bf Precision}& \textit{\bf Recall}& \textit{\boldmath{$F_{\textit{1}}$}\bf -score}\\
            \hline
            \textit{\toolname{}} & 0.8697 & 0.9016 & 0.8730\\
            \hline
            \textit{\toolname{} with only API} & 0.3408 & 0.3658 & 0.3408\\
            \textit{relevance classifier} &  & &\\
            \hline
            \textit{\toolname{} with only DATYS+} & 0.8620 & 0.8723 & 0.8530\\
            \hline
            \end{tabular}
        \label{table:varycomponent}
    \end{center}
\end{table}

Table~\ref{table:benchmarkTest} shows the performance of the DATYS and \toolname{} in finding threads that are relevant to the given API. \toolname{}, in general, outperforms DATYS. 
On average, \toolname{} achieves an F1-score of 0.873, which is an improvement of 13.9\% compared to DATYS. \toolname{} also beats DATYS in terms of precision and recall.

\subsection{RQ2: Ablation Study}\label{subsec:ablation}

\begin{table}[t]
    \caption{Average Precision, average Recall, and average \boldmath{$F_{\textit{1}}$}-score of testing sets when \combineVar{} varies}
    \begin{center}
        \begin{tabular}{|c|c|c|c|}
            \hline
            \textit{\combineVarMath{}}& \textit{Avg. Precision} & \textit{Avg. Recall} & \textit{Avg. \boldmath{$F_{\textit{1}}$}-score}\\
            \hline
            \textit{$0$} & 0.3420 & 0.3658 & 0.3408\\
            \hline
            \textit{$0.1$} & 0.4485 & 0.4653 & 0.4441\\
            \hline
            \textit{$0.2$} & 0.8650 & 0.8925 & 0.8641\\
            \hline
            \textit{$0.3$} & \textbf{0.8697} & \textbf{0.9016} & \textbf{0.8730}\\
            \hline
            \textit{$0.4$} & 0.8684 & 0.8934 & 0.8689\\
            \hline
            \textit{$0.5$} & 0.8588 & 0.8723 & 0.8097\\
            \hline
            \textit{$0.6$} & 0.8588 & 0.8723 & 0.8510\\
            \hline
            \textit{$0.7$} & 0.8606 & 0.8723 & 0.8521\\
            \hline
            \textit{$0.8$} & 0.8606 & 0.8723 & 0.8521\\
            \hline
            \textit{$0.9$} & 0.8606 & 0.8723 & 0.8521\\
            \hline
            \textit{$1.0$} & 0.8620 & 0.8723 & 0.8530\\
            \hline
            \end{tabular}
        \label{table:ablationTestSet}
    \end{center}
\end{table}

Table~\ref{table:varycomponent} shows how well each component in \toolname{} is. 
Since the  ``\TextCodeClassifier{}-only'' version of \toolname{} gives worst result, \TextCodeClassifier{} may not be able to resolve the task well independently. Partly, this might be be due to the limited amount of the training data (i.e., only 253 training threads). In addition, the ``DATYS+ only'' version of \toolname{} performs much better compared to the ``\TextCodeClassifier{}-only'' version. However, \toolname{} is still better than both of them. It demonstrates that both components are useful and essential.

\subsection {RQ3: Effect of the \combineVar{}}

Table~\ref{table:ablationTestSet} shows the performance of \toolname{} in the training set when we vary the values of \combineVar{}. The bold numbers in each row of the table are the average \fonescores{} of the chosen values of \combineVarMath{} in the training sets. Similarly, Table~\ref{table:fold1FindX} the performance of \toolname{} in the test set when we vary the values of \combineVar{}. The highest \fonescore{} for both the training and testing set is achieved when the value of the \combineVar{} is equal to 0.3. It demonstrates that our strategy to pick the value of the \combineVar{} that leads to the best performance in the training data works really well.

\begin{table}[t]
    \caption{Average Precision, average Recall, and average \boldmath{$F_{\textit{1}}$}-score of training sets when \combineVar{} varies}
    \begin{center}
        \begin{tabular}{|c|c|c|c|}
            \hline
            \textit{\combineVarMath{}}& \textit{Avg. Precision} & \textit{Avg. Recall} & \textit{Avg. \boldmath{$F_{\textit{1}}$}-score} \\
            \hline
            \textit{$0$} & 0.6565 & 0.6685 & 0.6473\\
            \hline
            \textit{$0.1$} & 0.7111 & 0.7272 & 0.7080\\
            \hline
            \textit{$0.2$} & 0.8261 & \textbf{0.8506} & 0.8269\\
            \hline
            \textit{$0.3$} & \textbf{0.8328} & 0.8498 & \textbf{0.8329}\\
            \hline
            \textit{$0.4$} & 0.8265 & 0.8410 & 0.8254\\
            \hline
            \textit{$0.5$} & 0.8159 & 0.8234 & 0.8097\\
            \hline
            \textit{$0.6$} & 0.8132 & 0.8234 & 0.8079\\
            \hline
            \textit{$0.7$} & 0.8132 & 0.8234 & 0.8079\\
            \hline
            \textit{$0.8$} & 0.8132 & 0.8234 & 0.8079\\
            \hline
            \textit{$0.9$} & 0.8132 & 0.8234 & 0.8079\\
            \hline
            \textit{$1.0$} & 0.8180 & 0.8191 & 0.8073\\
            \hline
            \end{tabular}
        \label{table:fold1FindX}
    \end{center}
\end{table}

\section{Discussion}\label{sec:discussion}

\subsection{Cases where \toolname{} outperforms DATYS}\label{subsec:outperformingCases}
\noindent {(1)\em\ The relevant thread does not contain the type name of the given API method}

\begin{figure}[t]
\centerline{\includegraphics[width=\columnwidth]{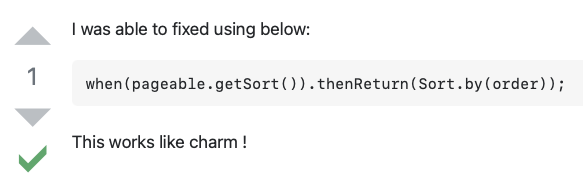}}
\caption{Thread 56135373 on Stack Overflow where API {\em } is referred by a code snippet of the thread}
\label{fig:discuss_1}
\end{figure}

\vspace{0.1cm}\noindent Figure~\ref{fig:discuss_1} shows the example of a case where the content of the thread does not relate to the given API method.
The thread contains paragraph and code snippet of a Stack Overflow thread with ID 56135373\footnote{https://stackoverflow.com/questions/56135373/}. {\em org.mockito.stubbing.OngoingStubbing.thenReturn}\footnote{\url{https://javadoc.io/doc/org.mockito/mockito-all/2.0.2-beta/org/mockito/stubbing/OngoingStubbing.html}} is the API method the thread refers to. 

From the content of the thread, it would be difficult to find the relevance between the text written in the paragraphs and the given API method (i.e., {\em \url{org.mockito.stubbing.OngoingStubbing.thenReturn}}) since the type (i.e., {\em OngoingStubbing}) does not appear in the thread.
The text only shows the user view towards the code snippet without having a description mentioning the application or usage of the observed API method invocation (e.g., {\em thenReturn } in the code snippet of Figure~\ref{fig:discuss_1}). Sentences such as {\em "This works like charm!"} do not provide much information to identify whether the observed API method refers to the given API. 

\begin{figure}[ht]

\begin{subfigure}{\linewidth}
  \centering
  \includegraphics[width=0.8\linewidth]{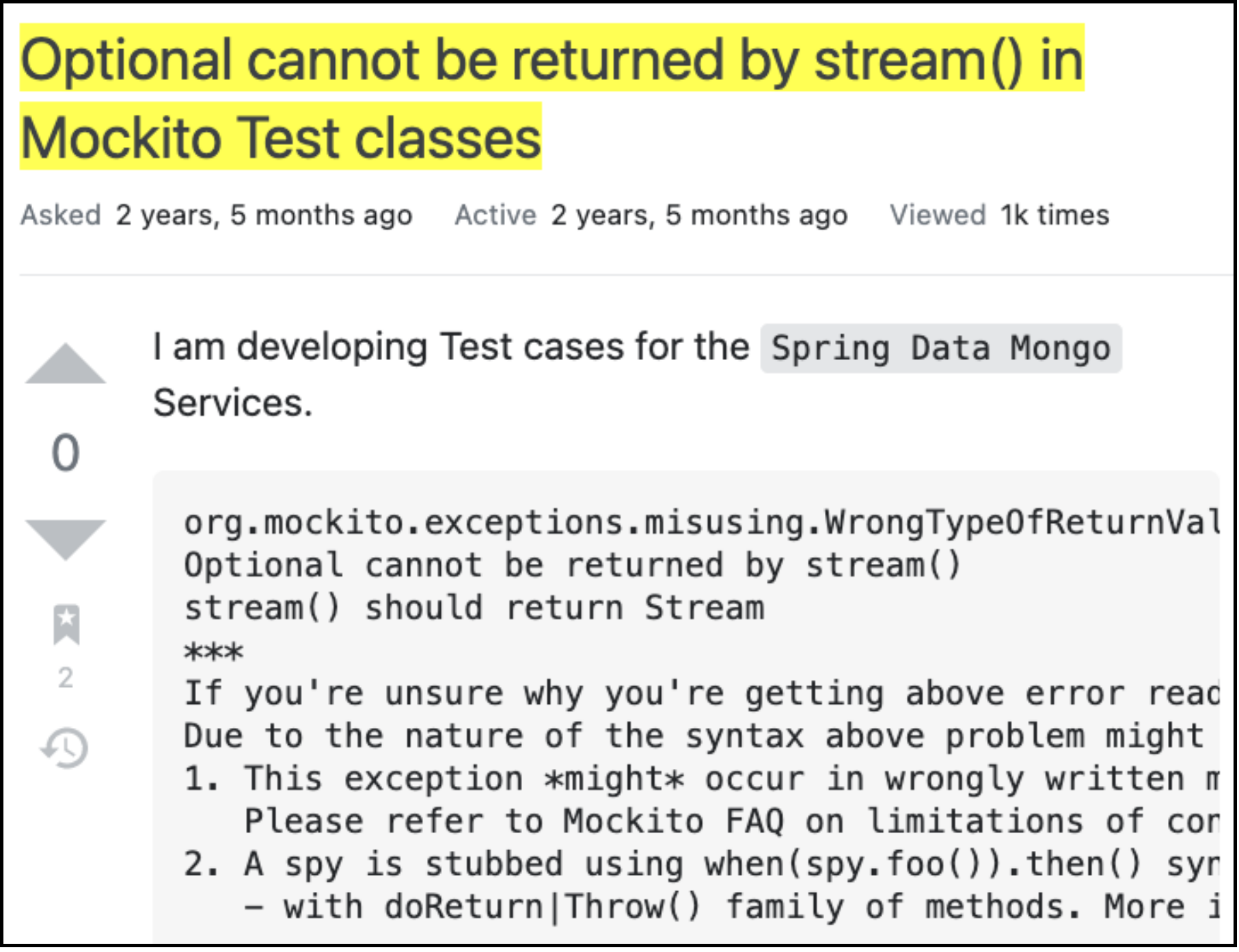}
  \caption{}
  \label{fig:56135373title}
\end{subfigure}
\begin{subfigure}{\linewidth}
  \centering
  \includegraphics[width=0.8\linewidth]{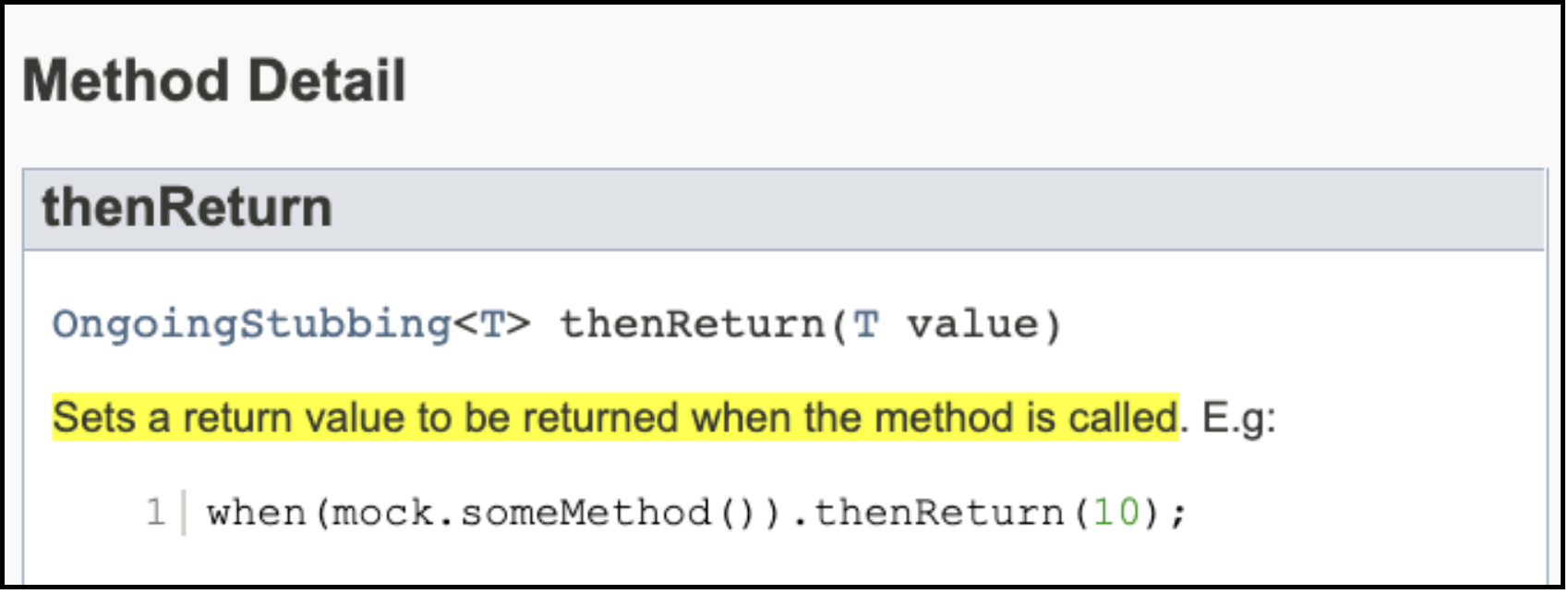}
  \caption{}
  \label{fig:56135373comment}
\end{subfigure}
\caption{The similarity in semantic meaning between the API comment of method {\em \url{org.mockito.stubbing.OngoingStubbing.thenReturn}} in Figure \ref{fig:56135373title} and the textual content (i.e., the title) of thread 56135373 in Figure~\ref{fig:56135373comment} }
\label{fig:semanticSimExample}
\end{figure}

Therefore, we leverage the content of the thread which might be relevant to the content of the API method. For example, in the thread above, its title which is shown in Figure~\ref{fig:56135373title}, {\em "Optional cannot be returned by stream() in Mockito Test classes"}, relates to the comment of the given API which is {\em Sets a return value to be returned when the method is called} in Figure~\ref{fig:56135373comment}. Due to this feature, FACOS can successfully consider this thread as relevant while DATYS missed it.
% \kien{Since both DATYS and \toolname{} consider the title as one of the content to perform the search task, DATYS cannot accurately find the relevance between the thread and the given API when the type of it (i.e., {\em OngoingStubbing}) does not exist in the thread. Thanks to the \TextCodeClassifier{} component, \toolname{} can successfully consider this thread as relevant while DATYS missed it.}

\vspace{0.2cm}\noindent {(2)\em\ The irrelevant thread contains the type name of the given API method}

\begin{figure}[t]
\centerline{\includegraphics[width=0.8\columnwidth]{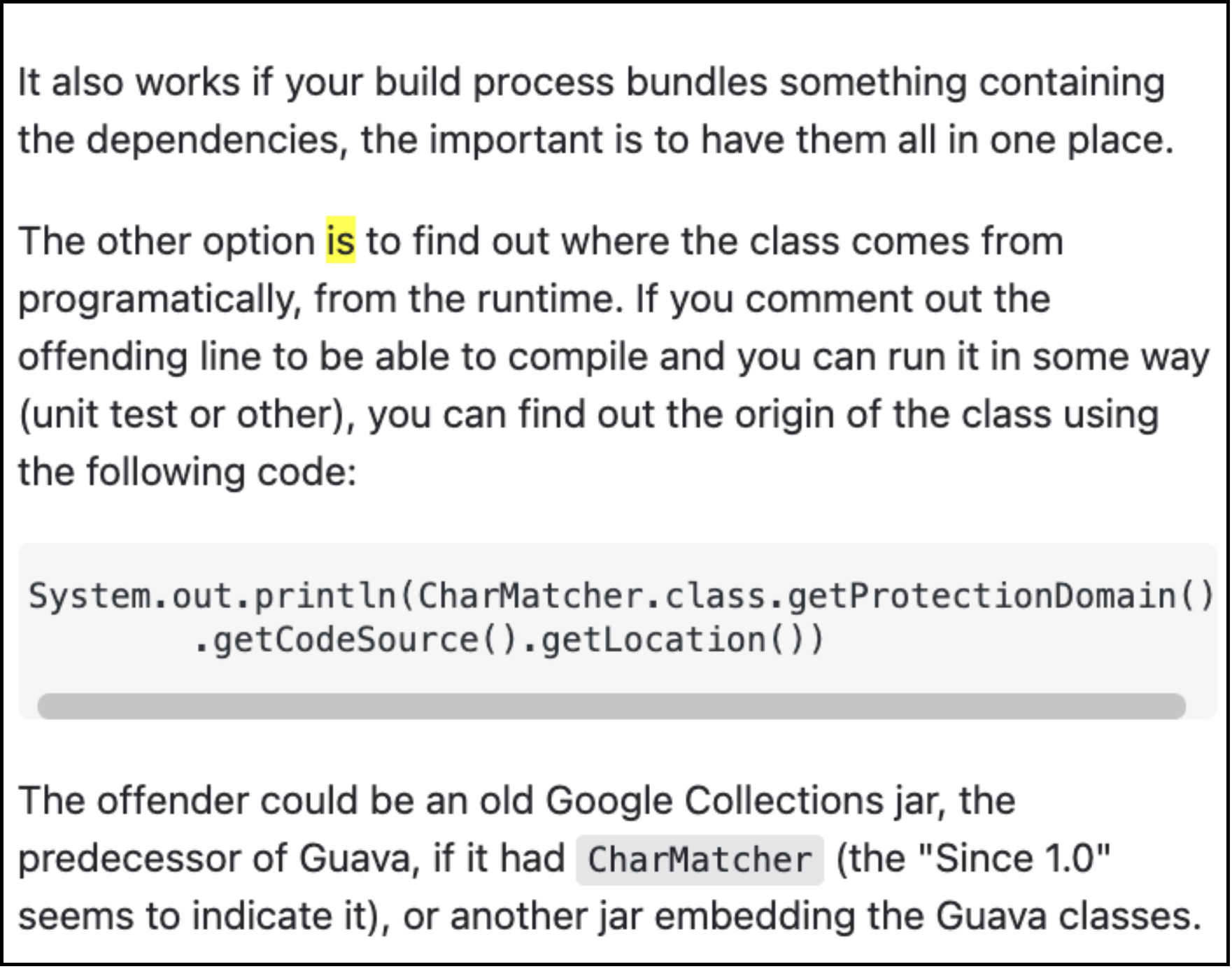}}
\caption{Thread 16919751 on Stack Overflow where API {\em \url{com.google.common.base.CharMatcher.is}} is not referred by content of the thread.}
\label{fig:discuss_2}
\end{figure}

\noindent
An example of this case is shown in Figure~\ref{fig:discuss_2}. In the thread\footnote{\url{https://stackoverflow.com/questions/16919751/}}, the given API method is {\em com.google.common.base.CharMatcher.is} and there is a word that matches the simple name of the API method {\em is} which we highlighted. Since the type of the given API method (e.g., {\em CharMatcher}) appears in both the textual content and the code snippet, DATYS mistakenly accepts the thread as referring to the given API. By leveraging the semantic knowledge learnt by the \TextCodeClassifier{}, \toolname{} is able to detect the irrelevance between the textual content, code snippet around the word {\em is} and the API comment and implementation code. \toolname{} can conclude that the thread is irrelevant to the given API {\em com.google.common.base.CharMatcher.is}.

\subsection{Case where \toolname{} fail to exclude irrelevant threads}
\noindent

\begin{figure}[t]
\centerline{\includegraphics[width=0.8\columnwidth]{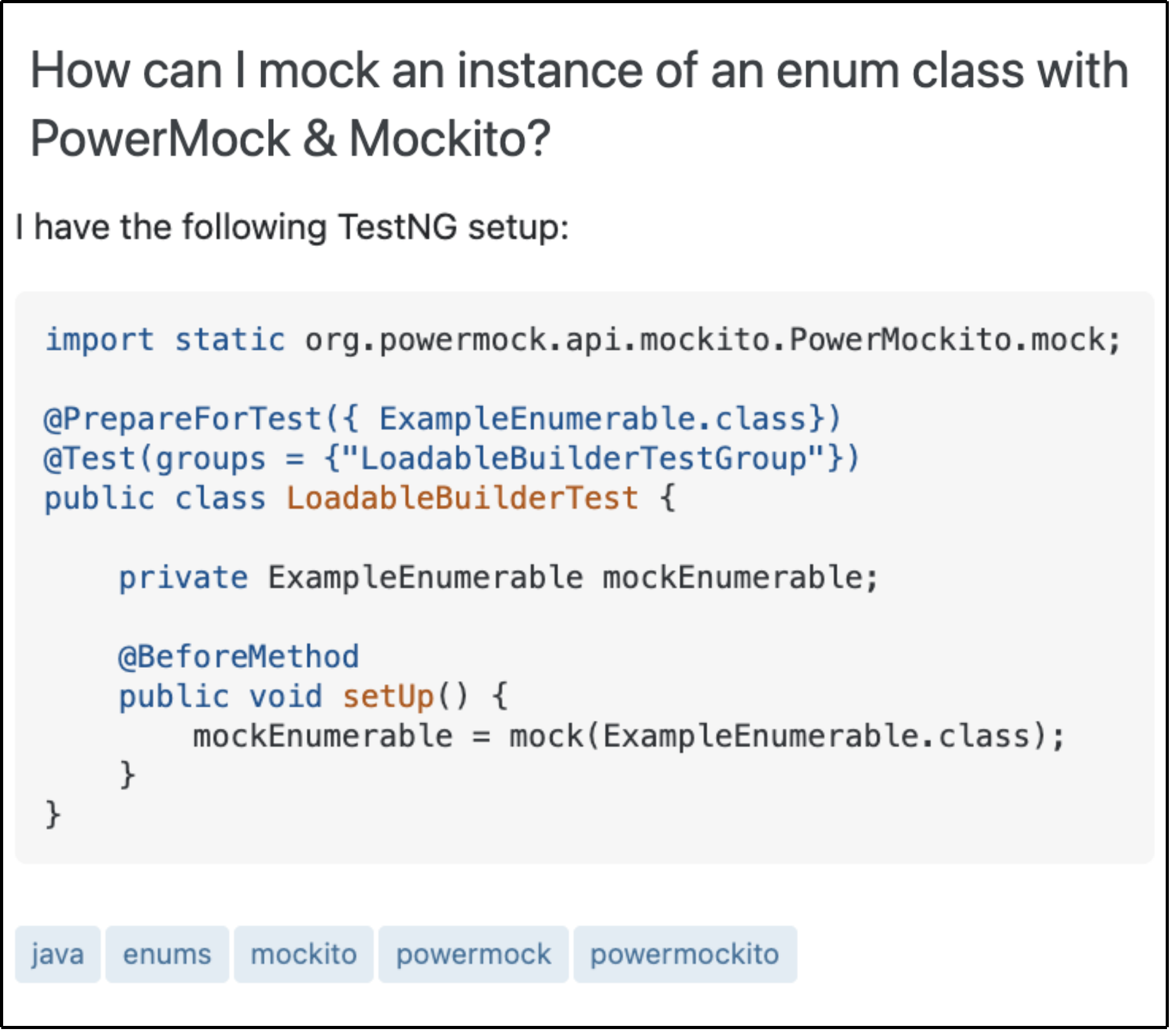}}
\caption{Thread 30127057 on Stack Overflow that \toolname{} falsely recognize as referring to the API method {\em \url{org.mockito.Mockito.mock}}.}
\label{fig:disscuss_4}
\end{figure}

Figure~\ref{fig:disscuss_4} shows a case where \toolname{} fail to exclude the thread\footnote{\url{https://stackoverflow.com/questions/30127057/}} out of the relevant results for the given API method {\em org.mockito.Mockito.mock}. The issue occurs when there is an API method that has a similar functionality as the given API method. These two methods usually have the same simple name and highly similar functionality description.

\begin{figure}[t]
\centerline{\includegraphics[width=0.8\columnwidth]{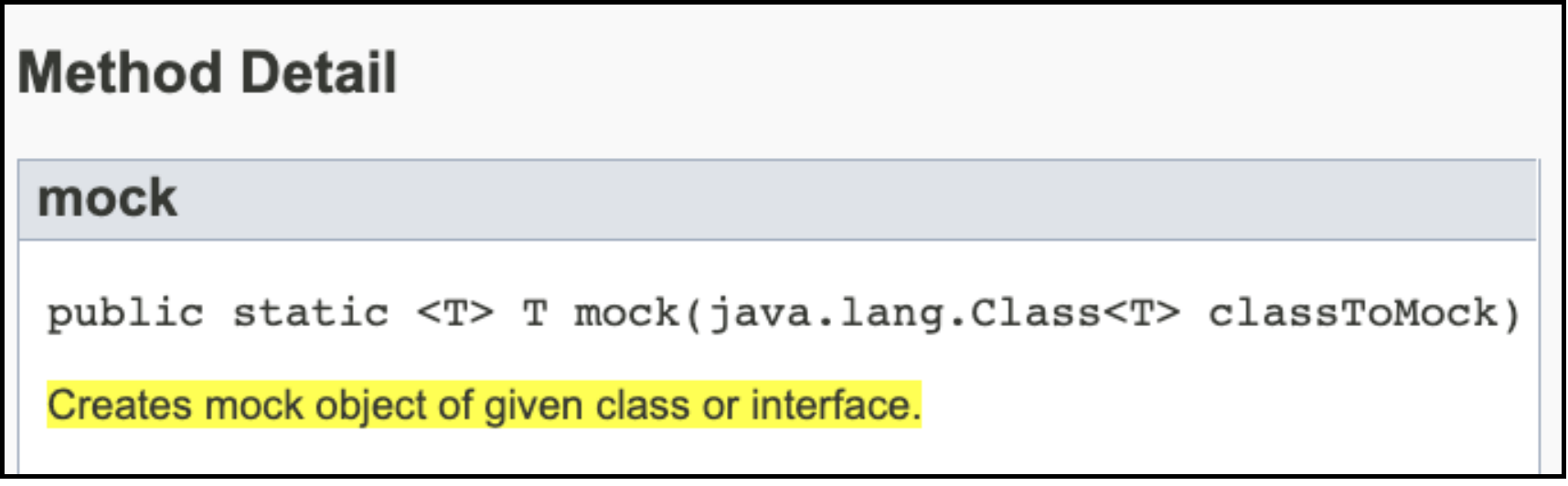}}
\caption{The API comment of method {\em \url{org.mockito.Mockito.mock}}}
\label{fig:disscuss_5}
\end{figure}

In Figure~\ref{fig:disscuss_4}, {\em PowerMock} and {\em Mockito}, perform similar functions such as mocking (i.e., creating a version of a service in order to quickly and reliably run tests on that service\footnote{https://circleci.com/blog/how-to-test-software-part-i-mocking-stubbing-and-contract-testing/}). 
Since both of them have the API method whose simple name is {\em mock}, and both their {\em mock} methods have the same API signature (i.e., parameters, return type),
it would be easy to mistakenly recognize one as the other, even for a human. Figure~\ref{fig:disscuss_5} shows the comment of API method {\em org.mockito.Mockito.mock}\footnote{\url{https://javadoc.io/static/org.mockito/mockito-all/2.0.2-beta/org/mockito/Mockito.html}}, which is {\em Creates mock object of given class or interface}. Because of the similarity between the API method from {\em Mockito} library and the title of thread 30127057 in Figure~\ref{fig:disscuss_4}, \toolname{} wrongly recognizes that the simple API name {\em mock} in the thread refers to the given API method {\em org.mockito.Mockito.mock}. In fact, the simple API name {\em mock} refers to the one from {\em PowerMock} library.

\subsection{Adding Semantic Information for API Content Search}
The API method and relevant information search does not always utilize syntactic information. For some libraries, some methods are chained (as in Figure \ref{fig:discuss_1}), so the type is not displayed. It requires a type system to determine its type. In addition, the Type system is not 100 percent reliable due to a lack of import information, variables, etc. 
It is ineffective to use the similarity of word representations provided by language models for API contents search. API contents search is unique to other tasks such as code search. In code search, due to the term "transfer" and "convert" being related to currency, the method named "transferSGDToUSD" might be the answer to the query "convert SGD to USD." However, searching for an API method requires explicit mention of the API name in the query, which could be difficult as different APIs may share the same API method names. To solve the problem, solely using a model that only learns the syntactic meaning of textual and code content is insufficient.
Therefore, we incorporated semantic information for API content search, which generated better results, as we have demonstrated from the results of our experiment.

\subsection{Threats to validity}\label{sec:threats_to_validity}
A threat to internal validity is related to experiment bias. We obtain our dataset from another work. We also run the baseline using the code provided by the author. We then check our code multiple times to ensure that we do not make mistakes. We believe there should be little threats to internal validity. We also release the dataset and the code for our experiments for all to use.

A threat to external validity is on whether the approach is applicable to other platform other than Stack Overflow.
This experiment mainly focuses on Stack Overflow and Java programming language, therefore, it is uncertain whether \toolname{} can be applied on other discussion venues that also talk about API issues. The potential platform might be Reddit, which has sub-reddit (i.e., a place gathering Reddit's threads discussing a particular problem) discussing programming languages and frameworks. It has title, textual and code content as same as Stack Overflow. Thus, the similarity suggests that we can potentially apply FACOS to Reddit too. We leave this possibility for a future work.
Regarding the threat that changing targeting programming language would affect the accuracy of \toolname{}, although we focus only on Java, the features (e.g., API comment/documentation, API implementation code, fully qualified name, class/type name, etc.) required for the approach can be found from other programming languages. Therefore, we leave this as a future work to study whether it may work well when applied to other programming languages.

There is a also threat of construct validity on whether precision, recall, and \fonescore{} is a suitable evaluation metric for our task. Our task is a classification task. Many work in software engineering has used precision, recall, and F1-score as the evaluation metric for classification task~\cite{kien2021datys, ye2018apireal, huang2018identifying,prana2019categorizing}. Thus, we believe the threats are minimal.

\section{Related Work}\label{sec:related_works}

\subsection{API Disambiguation}
Many past works~\cite{ye2018apireal,saifullah2019learning,antoniol2002recovering,marcus2003recovering,bacchelli2010linking,dagenais2012recovering,subramanian2014live,phan2018statistical} deals with API disambiguation. There are two main groups: informal text disambiguation~\cite{antoniol2002recovering,marcus2003recovering,bacchelli2010linking,dagenais2012recovering,ye2018apireal} and code snippet disambiguation~\cite{subramanian2014live,phan2018statistical,saifullah2019learning}. As suggested by its name, the first aims to disambiguate API mentions in textual content while the second deals with disambiguating API mention in code snippets.

For informal text disambiguation, several work utilizes classical information retrieval approaches such as Vector Space Model and Latent Semantic Indexing to disambiguate the API mentions~\cite{antoniol2002recovering,marcus2003recovering,bacchelli2010linking} while some others use  heuristics~\cite{dagenais2012recovering}.
Bacchelli et al~\cite{bacchelli2010linking} combined string matching and information retrieval algorithms to link emails to source code entities. 
Dagenais and Robillard~\cite{dagenais2012recovering} identified Java APIs mentioned in support channels (e.g., mailing list, forums), documents, and code snippets.
Ye \textit{et al.}~\cite{ye2018apireal} worked on API disambiguation in the textual content of Stack Overflow thread by utilizing mention-mention similarity, mention-entry similarity, and scope filter. Luong et al.\cite{kien2021datys} used type scoping to disambiguate the API mentions in Stack Overflow thread. 

The work on API disambiguation on Stack Overflow thread can be viewed as another side of the coin of the task in finding threads that are relevant to the API. When we disambiguate an API in a thread, the disambiguated API is relevant to the thread as the thread is talking about the API.

\subsection{API Resource Retrieval}
Several studies have explored how to search for the code for API and related information retrieval. Lv et al. \cite{Codehow} proposed Codehow to deal with the lack of query understanding ability of the existing tool. By expanding a user query with APIs, Codehow can identify potential APIs and perform a code search based on the Extended Boolean model, which considers the impact of APIs on code search. Gu et al. \cite{DeepAPI} proposed DeepAPI to search for API usage sequences. As opposed to assuming a bag of words, it learns the sequence of words within a query and the sequence of APIs associated with it. DeepAPI encodes a single user query into a fixed-length context vector to generate an API sequence.

Other studies have also exploited different aspects of APIs and natural language to better retrieve the APIs and their related information. The techniques include: using global and local contexts of the queries \cite{global_local_context}, leveraging usage similarity for effective retrieval of API examples \cite{usage_similarity}, employing word embeddings to document similarities for improved API retrieval \cite{word_embeddings}, exploiting user knowledge \cite{User_Knowledge}, and task-API knowledge gap \cite{TaskAPI} during retrieval of semantically annotated API operations.

Wang et al. \cite{wang2020trans} developed a transformer-based framework for unifying code summarization and code search. Shahbazi et al. \cite{shahbazi2021api2com} proposed API2Com to improve automatically generated code comments by fetching API documentations.
Alhamzeh et al. \cite{alhamzeh2021distilbert} built DistilBERT-based argumentation retrieval for answering comparative questions.
Dibia et al. \cite{dibia2020neuralqa} and Vale et al.\cite{vale2021towards} developed a usable library for question answering with contextual query expansion and a question-answering assistant for software development using a transformer-based language model, respectively.
Ciniselli et al.\cite{ciniselli2021empirical} performed an empirical study on the usage of Transformer Models for code search and completion.

Our study also work on API resource retrieval. Specifically, we retrieve Stack Overflow threads that are relevant to a target API that we are searching for.

\subsection{Contribution of StackOverflow for API documentation}

Treude et al. \cite{treude2016augmenting} studied the augmenting API documentation with insights from stack overflow.
\cite{parnin2012crowd} explored the Crowd documentation by examining the dynamics of API discussions on Stack Overflow, whereas \cite{squire2015should} dubbed the Stack Overflow as the Social Media for Developer Support in terms of provided utilities.
\cite{ahasanuzzaman2018classifying} and \cite{baltes2020contextual} worked on classifying stack overflow posts on API issues and contextual documentation referencing on stack overflow.
The dichotomy of these studies is notable where some research like \cite{uddin2015api} and \cite{zhang2018code} studies how API documentation fails via the API misuse on stack overflow, other studies \cite{meldrum2017crowdsourced, rocha2016automated} heavily lean on the Crowdsourced knowledge on stack overflow for automated API documentation with tutorials.
Similarly, crowdsourced knowledge was hailed by \cite{gomez2013study} and \cite{li2016discussion}, who explored the innovation diffusion and web resource recommendation for hyperlinks through link sharing on stack overflow.

Our work support the effort in this line of study. \toolname{} can automatically find threads about a particular API in Stack Overflow that can be augmented to the corresponding API documentation.

\subsection{Word Sense and Entity Disambiguation Study}
There are several works focused on disambiguation task~\cite{dagenais2012recovering,subramanian2014live,nguyen2018statistical,phan2018statistical}.
We also have found a variety of word sense and entity disambiguation methods employed for different objectives \cite{steiner2013adding, Karimzadeh_geoText, patwardhan2005senserelate, Jose_Tweet_disambiguate, foppiano2020entity, zwicklbauer2013we, mandalios2018geek, klein2002combining, chen2009fully}. These studies have solved myriads of problems via solving lexical disambiguations in literature. The task of word sense disambiguation is to identify a target word's intended meaning by examining its context. Researchers have used Word Sense Disambiguation to predict election results by enhanced sentiment analysis on Twitter data. Researchers have associated place-name mentions in unstructured text with their actual references in geographic space using word disambiguation. Other research has also proposed unsupervised, knowledge-Free, and interpretable Word Sense Disambiguation for various applications.
Researchers used this approach to add meaning to social network posts when it comes to named entity recognition and disambiguation. Different Entity-fishing tools were also developed for facilitating the recognition and disambiguation service. In recent years, tools that allow researchers to recognize and extract named entities have become increasingly popular.

\section{Conclusion and Future Work}\label{sec:conclusion}
We present \toolname{}, an approach to search Stack Overflow threads that refer to API of which users or tools may want to find the usage. We utilize the semantic and syntactic features of the paragraphs and code snippets in a thread to determine whether the thread is related to a given API. Our evaluation shows that \toolname{} has an improvement compared to DATYS when adapting both approaches to the search task. 
We have added a weight parameter to balance the usage of syntactic and semantic information for retrieving API mentions and related threads. We have proved the utility of the weight factor by incorporating an ablation study.
In future, we plan to improve our approach with larger dataset which has more threads and APIs. Also, we plan to make our approach become robust with more programming languages so that it can be more useful to developers.

\vspace{0.2cm}\noindent {\bf Replication Package.} The source code for \toolname{} is available at \url{https://anonymous.4open.science/r/facos-E5C6/}.

\balance
\bibliography{facos}
\bibliographystyle{IEEEtran}
\vspace{12pt}

\end{document}